\documentclass{svjour2}                    
\smartqed  
\usepackage{graphicx}
\usepackage{latexsym}
\usepackage{marvosym}
\usepackage{amsbsy}

\begin{document}

\title{Production of Slow Protonium in Vacuum
}


\author{N.~Zurlo         \and
        M.~Amoretti         \and
        C.~Amsler \and
        G.~Bonomi \and
C.~Carraro \and
C.L.~Cesar \and
M.~Charlton \and
M.~Doser \and
A.~Fontana \and
R.~Funakoshi \and
P.~Genova \and
R.S.~Hayano \and
L.~V.~J{\o}rgensen \and
A. Kellerbauer \and
V.~Lagomarsino \and
R. Landua \and
E.~Lodi~Rizzini \and
M.~Macr\`\i \and
N.~Madsen \and
G.~Manuzio \and
D.~Mitchard \and
P.~Montagna \and
L.G.~Posada \and
H.~Pruys \and
C.~Regenfus \and
A.~Rotondi \and
G.~Testera \and
D.P.~Van~der~Werf \and
A.~Variola \and
L.~Venturelli \lastand
Y.~Yamazaki
}

\authorrunning{N. Zurlo et al.} 

\institute{ N.~Zurlo ({\Letter}) \and E.~Lodi~Rizzini \and L.~Venturelli \at
Dipartimento di Chimica e Fisica per l'Ingegneria e per i Materiali, Universit\`a di Brescia, 25133 Brescia, Italy \\
Istituto Nazionale di Fisica Nucleare, Gruppo Collegato di Brescia, 25133 Brescia, Italy \\
              \email{zurlo@bs.infn.it}           
\and
            M.~Amoretti \and M.~Macr\`\i \and G.~Testera \and A.~Variola \at
Istituto Nazionale di Fisica Nucleare, Sezione di Genova, 16146 Genova, Italy
\and
            C.~Amsler \and H.~Pruys \and C.~Regenfus \at
Physik-Institut, Z\"urich University, 8057 Z\"urich, Switzerland
\and
            G.~Bonomi \at
Dipartimento di Ingegneria Meccanica, Universit\`a di Brescia, 25123 Brescia, Italy \\
Istituto Nazionale di Fisica Nucleare, Sezione di Pavia, 27100 Pavia, Italy
\and
          C.~Carraro \and V.~Lagomarsino \and G.~Manuzio \at
Dipartimento di Fisica, Universit\`a di Genova, 16146 Genova, Italy \\
Istituto Nazionale di Fisica Nucleare, Sezione di Genova, 16146 Genova, Italy
\and
          C.L.~Cesar \at
Instituto de Fisica, Universidade Federal do Rio de Janeiro, Rio de Janeiro 21945-970, Brazil 
\and
          M.~Charlton  \and D.~Mitchard \and L.~V.~J{\o}rgensen \and N.~Madsen \and D.P.~Van~der~Werf \at
Department of Physics, University of Wales Swansea, Swansea SA2 8PP, United Kingdom 
\and
           M.~Doser \and A. Kellerbauer \and R. Landua \at
Physics Department, CERN, 1211 Geneva 23, Switzerland \\
\emph{Present address} of A. Kellerbauer: Max Planck Institute for Nuclear Research, P.O.~Box 103980, 69029 Heidelberg, Germany.
\and
           A.~Fontana \and P.~Genova \and P.~Montagna \and A.~Rotondi \at
Dipartimento di Fisica Nucleare e Teorica, Universit\`a di Pavia, 27100 Pavia, Italy \\
Istituto Nazionale di Fisica Nucleare, Sezione di Pavia, 27100 Pavia, Italy
\and
           R.~Funakoshi \and R.S.~Hayano \and L.G.~Posada \at
Department of Physics, University of Tokyo, Tokyo 113-0033, Japan 
\and
           Y.~Yamazaki  \at
Atomic Physics Laboratory, RIKEN, Saitama 351-0198, Japan
}

\date{Received: date / Accepted: date}

\maketitle

\begin{abstract}
We describe how protonium, the quasi-stable antiproton-proton bound system,
 has been synthesized following the
interaction of antiprotons with the molecular ion H$_2^+$ in a nested Penning trap environment.
 From a careful analysis of the spatial
distributions of antiproton annihilation events in the ATHENA experiment,
evidence is presented for protonium production with sub-eV kinetic energies in states
around $n$ = 70, with low angular momenta.
This work provides a new 2-body system for study using laser spectroscopic techniques.

\end{abstract}

\section{Introduction}
\label{intro}
The availability of a high-quality low energy 
antiproton ($\bar \mathrm{p}$) beam delivered by 
the CERN Antiproton Decelerator (AD)
to the ATHENA, ATRAP and ASACUSA experiments has permitted the routine production of
stable pure antimatter systems (antihydrogen, $\bar \mathrm{H}$)
\cite{Amoretti:2002um,Gabrielse:2002} and
metastable mixed matter--antimatter systems (antiprotonic helium
or $\bar \mathrm{p}$He$^+$) \cite{Yamazaki:2002,Hori:2005},
both very important steps towards the goal of testing CPT symmetry.

Another metastable exotic atom that is also of great interest is antiprotonic
hydrogen ($\bar \mathrm{p}$p), also called protonium (Pn).
Its level structure is similar to that of hydrogen, but binding energies
are much larger and its simple two-body nature allows
an independent CPT test.
In fact, spectroscopic measurements on Pn,
made in near--vacuum conditions with high-precision laser techniques,
would make it possible to improve
the precision of the ``(anti)protonic Rydberg'' constant
or, equivalently, the (anti)proton to electron mass ratio.

Even though Pn has been studied in the past
(see e.g. \cite{Batty:1989,Montanet:2001} ),
in this paper we report a radically new method to produce Pn
resulting in emission with very low kinetic energy (from some meV to $\sim1$ eV)
in vacuum conditions that could open the way to laser spectroscopic studies.
Previous experiments produced Pn by injecting antiprotons into a molecular
hydrogen target (H$_2$), either liquid or gaseous. This results in a Pn lifetime which 
depends strongly upon the target density due to the effect of collisional de-excitation
and which makes spectroscopy impossible.
In the ATHENA apparatus Pn has been produced after a ``chemical''
reaction between $\bar \mathrm{p}$'s and molecular hydrogen ions (H$_2^+$)
trapped together with positrons ($e^+$) in a nested Penning trap.
Protonium production accompanies  the $\bar \mathrm{H}$ production
described in \cite{Amoretti:2002um}, and here we describe how the two modes
of  $\bar \mathrm{p}$ annihilation have been distinguished
(see also \cite{zurlo:???}).

\section{Experimental details}
\label{sec:2}
The ATHENA apparatus, described extensively in \cite{jorgensen:nima},
consisted of a multi-electrode system of cylindrical Penning traps,
2.5 cm in diameter and $\sim~1$~m in length kept
in an axial magnetic field of 3 T. 
In the 15 K cryogenic environment of the trap,
only hydrogen and helium were present in gaseous form, giving a
residual pressure of $\sim 10^{-12}$ Torr.
Antiprotons from the AD were caught,
cooled by electrons and stored
in the so--called mixing trap. The latter is 
a nested Penning trap, approximately  10 cm long,
that allowed  $e^+$s and $\bar \mathrm{p}$'s  to be 
confined simultaneously. Under typical conditions
the trap contained a spheroidal plasma of $\sim 3.5
\times 10^7~e^+$ and $\sim10^4$~$\bar \mathrm{p}$'s. 
The resulting $\bar p$ annihilations
were monitored for $\sim60$~s by detectors \cite{Regenfus:detector}
that, recording the passage of the charged pions,  
allowed  reconstruction of  annihilation vertices 
with an uncertainty of a few mm. 

In typical operating  conditions, annhilation of $\bar p$'s originate from:
\begin{itemize} 
\item $\bar \mathrm{H}$ formation followed by annihilation
on the electrode surface \cite{Amoretti:2002um,Niels:spatial_distribution}; 
\item  $\bar p$ annihilation in some well--defined 
``spots'' on the electrode walls due to radial transport
 \cite{Fujiwara:2004ra}; 
\item annihilation following interactions with residual gas atoms
or ions present in the trap.
\end{itemize}

\begin{figure}
  \includegraphics[width=\textwidth]{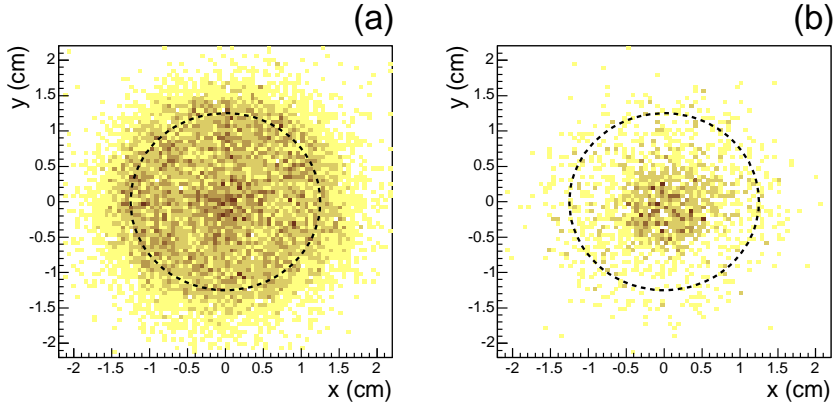}
\caption{Scatter plot of annihilation vertices
 projected on a plane perpendicular to the trap axis for
(a) cold mixing, (b) hot mixing. The dashed line represents
the trap wall.  }
\label{fig:1}       
\end{figure}

It was shown in \cite{Amoretti:2004ci} that, when the $e^+$ cloud was
kept at the trap environment cryogenic temperature of $\sim15$~K 
(a situation called ``cold mixing'' hereafter),
annihilations were mainly due to $\bar \mathrm{H}$,
even if some annihilations near the trap axis were present
(see also Fig.\ref{fig:1}a).
On the contrary, when the $e^+$ cloud was heated (by 
a radio--frequency drive applied to an electrode of the trap 
\cite{amoretti:modi,amoretti:modii}) to a temperature,
$T_e$, of several thousand K (about 8000~K for the data reported here,
which we call ``hot mixing'' hereafter) 
$\bar \mathrm{H}$ formation was
strongly  suppressed \cite{germano:temperatura}
and $\bar p$s annihilated mainly without forming $\bar \mathrm{H}$
(see also Fig.\ref{fig:1}b).

In the following, we will study in detail the latter  annihilations.

\section{Results and discussion}
\label{sec:3}
\begin{figure}
  \includegraphics[width=\textwidth]{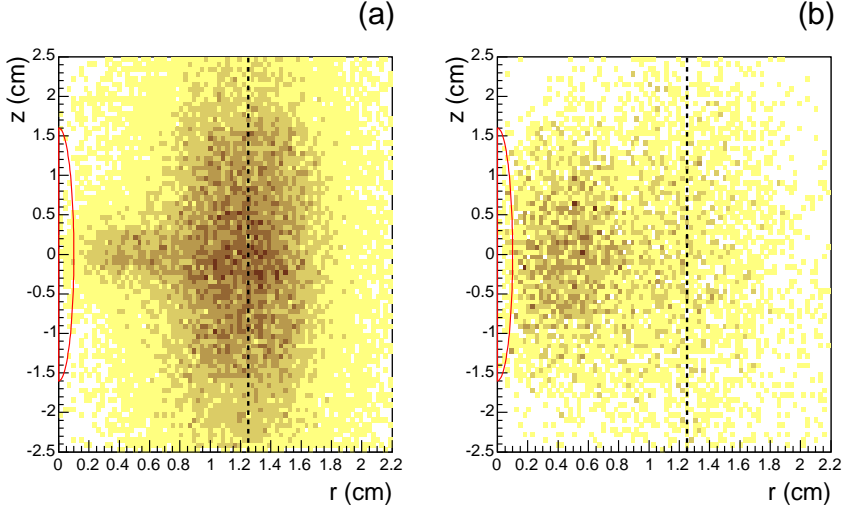}
\caption{Scatter plot of annihilation vertices
as a function of $r$ and $z$ for
(a) cold mixing, (b) hot mixing. The dashed line represents
the trap wall. The semi-ellipse indicates the size and position
of the positron cloud.  }
\label{fig:2}
\end{figure}

In Fig.\ref{fig:1}, we report the $x$--$y$ scatter plots (both coordinates
are in the plane perpendicular to the trap axis) for cold
and hot mixing data. In Fig.\ref{fig:2}, the corresponding
$r$--$z$ scatter plots are reported
 (where $r=\sqrt{x^2+y^2}$ is the radial position,
i.e. the distance from the trap axis, and $z$ is the axial coordinate,
measured from the symmetry plane of the trap).

\begin{figure}
  \includegraphics[width=\textwidth]{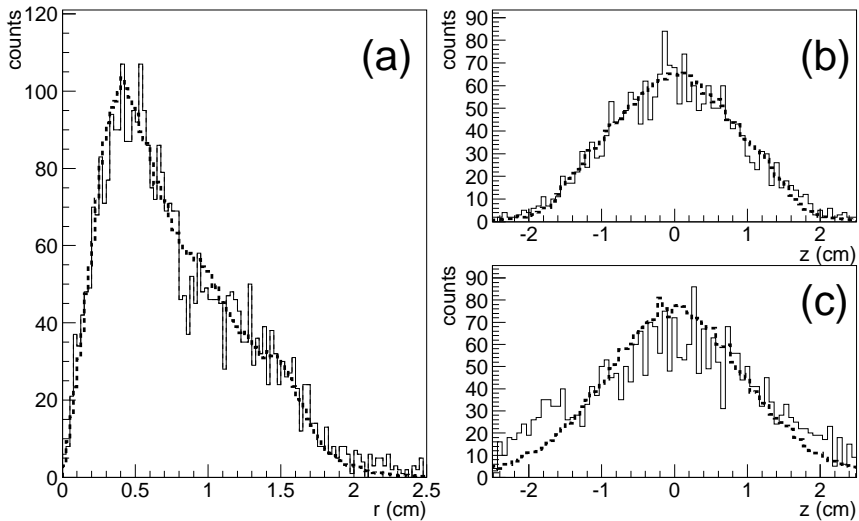}
\caption{Distributions of annihilation vertices for hot mixing 
(continuous line), with superimposed Monte Carlo simulation results
(dashed line).
(a) radial distribution; (b) axial distribution, for events near the trap
axis ($r < 0.5$ cm); (c) axial distribution, for events near the trap wall
($r>1$ cm). }
\label{fig:3}
\end{figure}

\begin{figure}
  \includegraphics[width=\textwidth]{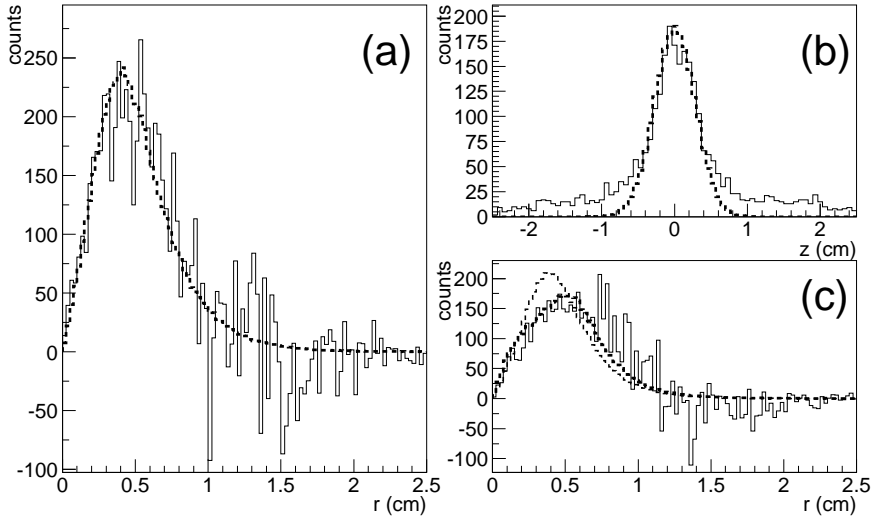}
\caption{Distributions of annihilation vertices for cold mixing,
(continuous line), with superimposed Monte Carlo simulation results
(dashed line).
(a) radial distribution, after removing the $\bar \mathrm{H}$ contribution
as discussed in the text; (b) axial distribution, for events near the trap
axis ($r < 0.5$ cm); (c) radial distribution, for the cold mixing sample
with wider positron plasma (in this  case, the thick dashed line
corresponds to Monte Carlo results with the correct  plasma parameters,
while the thin dashed line corresponds to Monte Carlo results with the
parameters of the narrower positron plasma; see text). }
\label{fig:4}
\end{figure}

Even though all the distributions are broadened by the uncertainty
in the vertex reconstruction, it is clear that two different
structures are merged in cold mixing case, while in hot mixing
just one appears.
For the cold mixing, besides the annihilations on the trap wall
situated at $r=1.25$ cm due mainly to $\bar \mathrm{H}$
(as shown in \cite{Amoretti:2004ci}) and having a relatively wider $z$-distribution
\cite{Niels:spatial_distribution}, there are
some annihilations situated at smaller $r$ and with a very
sharp $z$-distribution (see also Fig.\ref{fig:4}b).
 For hot mixing only the latter
are present, though their axial distribution is broader, 
as is the radial distribution   
(see also Fig.\ref{fig:3}a,b).

The capability of the ATHENA detector to detect the spatial and temporal
coincidence between $\bar p$ and $e^+$ annihilations,  and  therefore
to separate  $\bar \mathrm{H}$ from other annihilations, allows us to infer that 
for hot mixing almost all annihilations (even near the wall)
are not due to $\bar \mathrm{H}$,
while for cold mixing we have to distinguish between annihilations
near the trap axis, that are  not due to $\bar \mathrm{H}$ (apart from
a few poorly  reconstructed vertices), and near the wall, where the 
number of non-$\bar \mathrm{H}$ annihilations is negligible.

\begin{figure}
  \includegraphics[width=\textwidth]{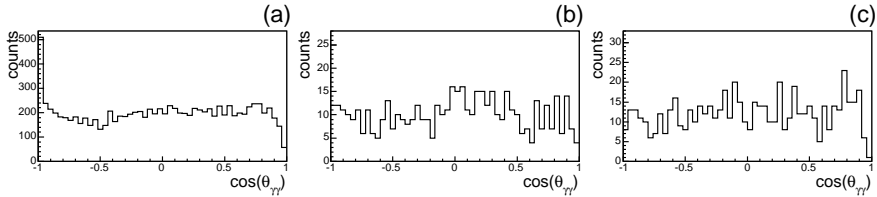}
\caption{Histogram of the cosine of the angle between the two detected photons
 ($\cos(\theta_{\gamma\gamma}$); see text) for
(a) cold mixing, on the wall ($r>1$ cm); (b) cold mixing, near the trap axis ($r<0.5$~cm);
(c) hot mixing.}
\label{fig:new}       
\end{figure}

This is clear if we look at Fig.\ref{fig:new}, where we consider annihilations
happening in coincidence with two (and only two)  photons detected,
and we plot the  cosine of the angle 
between the two detected photons, $\cos(\theta_{\gamma\gamma}$). This distribution
should have a peak in $\cos(\theta_{\gamma\gamma})=-1$ if annihilations
are  due to $\bar \mathrm{H}$, because of 
the two back--to--back 511 keV photons produced by the $e^+$ annihilation,
while we do not expect any peaks for annihilations not related to $\bar \mathrm{H}$. 
The former is the case for cold mixing on the wall (Fig.\ref{fig:new}a), whilst
the latter is the case for cold mixing near the trap axis (Fig.\ref{fig:new}b)
 and for  hot mixing (Fig.\ref{fig:new}c).

It is notable that the axial extent of the annihilation distributions is small compared 
with the size  of the nested trap, so they cannot be in-flight
annihilations of $\bar p$'s on residual gas.
However, their radial extent 
is too large to be explained by in-flight annihilations
on positive ions trapped inside the positron well.
So, only a process 
involving $\bar p$ interaction with an ion
to form a neutral system that can 
annihilate in flight, perhaps after leaving the trap,
 can explain the data.

Time-of-flight measurements following charged particle ejection from the trap
 excluded the presence of protons and of ions
of atoms more massive than helium.
However, reactions of $\bar p$ with helium ions would give distributions
that do not match the data, since 
 $\bar p$ He$^{++}$ 
gives rise to a charged system followed by rapid  annihilation,
and also for $\bar p$ He$^+$ the residual electron would be ejected in less than
10 ns \cite{korobov:2003}, again giving rise to a charged system.

Further important information on the $\bar p$--system formed can be found by 
exploiting the number of tracks  from each annhiliation vertex,
corresponding to the number of charged pions
produced during the annihilation, since this  depends on the particle 
with  which the $\bar p$ annihilates (see e.g. \cite{balestra:pbar-Ne:1989}).
Tab.\ref{tab:1} shows the ratios, $R_{23}$, of the number
of the reconstructed annihilation vertices having two tracks
to those with three tracks, for different data samples. 
In order to have a better understanding of these data, a Monte Carlo
simulation of $\bar p p$ (i.e. Pn) annihilations inside the ATHENA apparatus
has been performed. 
 Comparing these data, we see that annihilations on wall
differ from those originating 
inside the trap, which are compatible with $\bar p p$
annihilations both for cold mixing and hot mixing.

\begin{table}
\caption{Experimental and Monte Carlo results for the number of
charged pion tracks due to $\bar p$ annihilations. }
\label{tab:1}       
\begin{tabular}{lll}
\hline\noalign{\smallskip}
Data set & Ratio $R_{23}$ on wall& Ratio $R_{23}$ at centre\tabularnewline
  \\
\noalign{\smallskip}\hline\noalign{\smallskip}
Cold mixing& 1.35$\pm$0.01& 1.22$\pm$0.04 \\
Hot mixing& 1.38$\pm$0.10& 1.17$\pm$0.04 \\
{$\bar p$}s  only (no mixing) & 1.40$\pm$0.03&  \\
\noalign{\smallskip}
 Monte Carlo $\bar pp$&\multicolumn{2}{c}{1.19$\pm$0.01} \\
\noalign{\smallskip}\hline
\end{tabular}
\end{table}

Combining this information, we infer
 that the most probable $\bar p$--ion
reaction 
is:
\begin{equation}
\label{eq:1}
\bar{p}+ \mathrm{H^+_2}\rightarrow  {\mathrm {Pn}}(n,l) +\mathrm{H}.
\end{equation}
where the  Pn leaves the trap 
(because it is neutral),
and has an exponentially distributed annihilation lifetime (depending on  $n$ and $l$).
For relatively high $n$, Pn is formed in
a metastable state and its lifetime can be greater than $1 \mu$s
\cite{hayano:protonium_lifetime:1999}.

The H$_2^+$ ions required for reaction (\ref{eq:1})
 may have been created during the positron loading procedure \cite{jorgensen},
due to collisions with the residual H$_2$ gas, or during $\bar p$ loading
(similarly to that reported in \cite{Gabrielse:1999}). 
Measurements of charge during the dump of the trapped particles has indicated
that the number of ions could be as high as $10^5$, depending on the vacuum conditions.

In order to check our hypothesis, we performed a self-standing
Monte Carlo simulation
 to reproduce the observed annihilation distributions for hot mixing 
and cold mixing in the following way:
\begin{itemize}
\item we used the information we have on the $e^+$ plasma shape  \cite{amoretti:modi,amoretti:modii}
to generate the Pn starting point distributions (in particular,
it was found  that the $e^+$ plasma was approximately
a spheroid with radius $r_p$=1 mm and axial half-length $z_p$=16 mm,
rotating with a frequency of 300 kHz; i.e. a surface velocity of
about 2000 ms$^{-1}$).
\item assuming that Pn is produced in thermal equilibrium with 
the $e^+$ plasma, and knowing that Pn must inherit the 
drift velocity
from its charged components, we generated
a velocity distribution summing the thermal Maxwellian isotropic velocity
(with a mean value fixed by the measured temperature of the $e^+$ plasma)
with a velocity along the tangential direction 
(that is the same as inferred by the measured parameters
of the $e^+$ plasma, since the drift doesn't depend on either  the mass
or the charge of the particle).
\end{itemize}

Starting  with hot mixing (8000 K), the simulation was performed 
assuming that the Pn was produced on the surface of the  $e^+$ spheroid
(characterized by the parameters above),
with a Gaussian distribution along $z$
(having $\sigma$~=~10~mm and being limited to $|z|<z_p$)
or alternatively inside the $e^+$ spheroid with a uniform density.
The result is nearly independent of the assumed distribution
of starting positions, since the system is dominated by the thermal velocity
of 5600 ms$^{-1}$.
The simulated radial and axial distributions for the best fitted mean lifetime
($1.1 \mu$s)  are superimposed on the experimental data 
in Fig.\ref{fig:3}. The agreement is good, and in particular
the simulation shows that about 25\% of the Pn produced  reaches
the wall, as observed, and it also predicts that the axial distribution
 near the wall is slightly wider than that  near the axis
 (Fig.\ref{fig:3}b,c) due to the essentially isotropic spreading.

For cold mixing the analysis is  less straightforward,
because  the $\bar \mathrm{H}$  contribution
on the trap wall must be subtracted.
To do so, we considered the difference between a radial distribution
taken in a $z$-slice where Pn is present (e.g. $|z|<0.5$),
and 
one where there is no Pn  (e.g. 0.5 cm $<|z|<$ 1.5 cm), normalized on the tail for
$r$ $>$ 1.5 cm, which is essentially 
$\bar \mathrm{H}$ only. The result is shown in Fig.\ref{fig:4}a.
For the axial distribution,  isolating the  Pn signal 
is much simpler, because 
only annihilations near the axis of the trap need be considered (Fig.\ref{fig:4}b).

In order to simulate the cold mixing distributions,
the parameters ($r_p$, $z_p$, frequency rotation)
of the $e^+$ plasma were the same as for hot mixing,
and we generated Pn with a thermal velocity corresponding to 15 K 
(250 ms$^{-1}$ along each direction). 
However, in this case it is necessary to assume that Pn is produced in a very narrow region
around $r=r_p$ and $z=0$ (we generated it with a Gaussian distribution
along $z$ with $\sigma=2.5$ mm).
The best fitted   mean lifetime is 1.1 $\mu$s, as found for hot mixing.

The agreement between the experimental and Monte Carlo data is good
(Fig.\ref{fig:4}a,b). Furthermore, Monte Carlo results show that
less than 0.5\% of Pn reached  the wall in this case,
thus confirming the consistency of our normalization technique.

For cold mixing, we were also able  to analyze another data sample, where
the positron plasma had different parameters ($r_p$=2.5 mm,  $z_p$=18 mm,
rotation  frequency of 80 kHz, surface velocity of
about 1300 ms$^{-1}$). 
Following the same prescription, we have  reproduced  the 
radial distribution of this sample (Fig.\ref{fig:4}c), showing 
there is a notable difference between the distribution of Pn annihilations
in this case and that for a narrower $e^+$ plasma.

It has to be stressed that it is impossible to reproduce  the experimental results
unless Pn is produced in a narrow region straddling the ``equator''
of the $e^+$ spheroid.
A possible and quite straightforward way to explain this, 
is to assume that there is some kind of (partial) separation between the 
$e^+$ and the H$_2^+$ ions, as expected  in case of thermal equilibrium
(see \cite{Dubin:1999,Oneil:1981}), and as has been observed for
a mixture of $e^+$ and $^9$Be$^+$ in \cite{jelenkovic:2003,jelenkovic:2002}.
In fact, for our experimental conditions,
assuming thermal equilibrium, H$_2^+$ ions experience
a centrifugal potential barrier of the order of 10 meV
meaning that, for 15 K, their thermal energy  ($\sim 1$ meV) is not enough
to allow the ions to penetrate the $e^+$ plasma. However, for  8000~K, the barrier
is negligible compared to the thermal energy  ($\sim 700$ meV) of the ions such that
they will be distributed uniformly inside the plasma.

Another constraint from our Monte Carlo simulation relates to the Pn kinetic energy.
In fact, the hot mixing and the cold mixing data cannot be reproduced 
using the same mean lifetime for both of them
if the system  has a recoil kinetic energy of the order of 1 eV or greater.
This probably means that the $\bar p$-H$_2^+$ collision 
happens preferentially as a resonant transfer,
 i.e. the dissociation energy of H$_2^+$ is taken care of
by the binding energy of Pn, so that its energy level, $n$, 
should be around 70.

Because we find  a mean lifetime of about $1.1 \mu$s, it can be inferred
\cite{hayano:protonium_lifetime:1999}  that the orbital
angular momentum, $l$, should be around 10. This could be a consequence of
the fact that, with such a slow relative collisional velocity,
the H$_2^+$ molecular ion will be strongly polarized, giving rise to
an almost collinear collision.

\section{Conclusions}
\label{sec:4}
In this paper, we have presented evidence for
the production of protonium in vacuum. It is formed in a metastable state
(with a lifetime of about $1.1 \mu$s, independent of the environment
temperature) and with near-thermal kinetic energies
(varying from some meV to less than 1~eV).
The number of produced protonium atoms was about 100 for each mixing cycle
($\sim 100$~s),
in which around $10^4$ $\bar p$'s were injected into the mixing trap,
while the estimated number of ions trapped with the positrons 
was typically $10^4-10^5$.
 Taking into account the recently achieved
capability of accumulating $\sim 10^8$~H$_2^+$
and storing $\sim$~5~$\times 10^6$ $\bar p$'s in some minutes
\cite{Yamazaki:2005,Oshima:2004},
our result opens up the possibility of performing detailed 
spectroscopic measurements on protonium as a probe of fundamental
 constants and symmetries.


%

\begin{acknowledgements}
This work was supported by CNPq, FAPERJ, CCMN/UFRJ (Brazil),
INFN (Italy), MEXT (Japan), SNF (Switzerland), SNF (Denmark) and EPSRC (UK).

\end{acknowledgements}



\end{document}